\documentclass[aps,pre,amsmath,amssymb]{revtex4}

\usepackage{graphicx}
\usepackage{amsmath}
\usepackage{verbatim}
\usepackage{xcolor} 

\usepackage{algorithmic}
\usepackage{algorithm}

\DeclareMathOperator{\Op}{Op}
\DeclareMathOperator{\sech}{sech}

\begin{document}
\title{Transparent boundary conditions for the sine-Gordon equation:\\ Modeling the reflectionless 
propagation of kink solitons on a line}
\author{K.K. Sabirov$^{1,5}$, J.R. Yusupov$^{2}$, M. Ehrhardt$^{3}$ and D.U. Matrasulov$^4$}
\affiliation{$^1$Tashkent University of Information Technologies, 108 Amir Temur Str., 100200, Tashkent Uzbekistan\\
$^2$Yeoju Technical Institute in Tashkent, 156 Usman Nasyr Str., 100121, Tashkent, Uzbekistan\\
$^3$Bergische Universit\"at Wuppertal, Gau{\ss}strasse 20, D-42119 Wuppertal, Germany\\
$^4$Turin Polytechnic University in Tashkent, 17
Niyazov Str., 100095, Tashkent, Uzbekistan\\
$^5$Tashkent State Technical University named after Islam Karimov, 2 Universitet Str., 100095, Tashkent, Uzbekistan}

\begin{abstract}
We consider the reflectionless transport of sine-Gordon solitons
on a line. Transparent boun\-dary conditions for the sine-Gordon
equation on a line are derived using the so-called potential approach.
Our numerical implementation of these novel boundary conditions proves the absence
of the backscattering in transmission of sine-Gordon solitons
through the boundary of the considered finite domains.
\end{abstract}

\maketitle

\section{Introduction}

Sine-Gordon solitons are an important class of nonlinear waves appearing in different branches of science and technology, e.g.\ propagation of fluxons in Josephson junctions in semiconductors, solids, DNA and tectonic plates (see,
e.g., the Refs.~\cite{Panos,Kivshar1,drazin,peyrard,ibach,ablowitz,Baron,Baron71,McCann,Bykov1,Bykov2,Yamosa,Yakushevich}, for review). Additionally, the sine-Gordon equation (SGE) appears as the continuous limit of the discrete sine-Gordon equation for the lattice wave field in the Frenkel-Kontorova (FK) model, a model of the dynamic behaviour of crystal defects in solid state. 
A spatially discrete SGE models the chain of point-like discrete Josephson junctions 
\cite{Sodano3,Sodano1,Sodano4,Sodano2}. Each spatial discretization corresponds to a different model.

Unlike other types of nonlinear waves, 
sine-Gordon solitons are relativistic and, hence, Lorentz invariant waves described by nonlinear partial differential equation  involving the d'Alembert operator $\displaystyle\Box=\partial_t^2-\partial_x^2$ (which is invariant under Lorentz transformations) and the sine of the unknown function. A remarkable feature of the sine-Gordon equation on a line is its integrability
and the admission of soliton (kink, antikink, breather, etc.) solutions \cite{Panos,Kivshar1}. 
So far, many aspects of mathematical and physical properties of the sine-Gordon equation and its soliton solutions  
have been extensively studied, both for traveling and standing waves. 
Recently, soliton dynamics in networks described in terms of sine-Gordon equation on metric graphs attracted some attention \cite{Hadi1,Hadi2,SGEEPL,SSGEPLA}. Utilization of such approach makes possible modeling the charged solitons in conducting polymers \cite{Chsol} and static solitons in branched Josephson junctions \cite{BJJEPL}.
However, despite the great progress made on this topic, some issues are still remaining unresolved.

This concerns, e.g., so-called transparent and absorbing boundary conditions
for one- and multi-dimensional sine-Gordon equations. Such boundary conditions are determined 
as those, which make equivalent (similar) the solution of a PDE on a given bounded domain to that in a 
whole space, so that no back scattering is possible at the boundary for incoming (outgoing) travelling waves.
In other words, the wave passing through the boundary does not ``feel'' it. 
So far, transparent boundary conditions have been studied for different wave equations 
having broad applications in physics, such as linear \cite{Arnold1998,Ehrhardt1999,Ehrhardt2001} and nonlinear \cite{Antoine,Matthias2008} Schr\"odinger, Dirac \cite{Hammer2014}, diffusion \cite{Wu} and Bogoliubov de Gennes \cite{Schwendt} equations. 
Recently, the concept of transparent boundary conditions have been extended to linear \cite{Jambul,Jambul02,Exciton}, nonlinear \cite{Jambul1} Schr\"odinger and Dirac \cite{Jambul2} equations on metric graphs.

Until today many different numerical schemes like compact schemes~\cite{Cui09,Cui10, Sari11},  predictor-corrector schemes~\cite{Cui09, Khaliq00}, 
energy-conservative finite difference schemes~\cite{Ben-Yu86, Fei94},  
Lattice-Boltzmann methods~\cite{Lai11}, radial basis functions~\cite{Dehgan08}, 
etc.\ were designed to solve numerically the sine-Gordon equation on the real line.
The authors simply considered a sufficiently large domain and supplied homogeneous Dirichlet or Neumann boundary condition.
Doing so they bypass the main challenge of this problem, namely how to treat appropriately the unbounded domain, since it is not clear what is `sufficiently large' and how does the simple chosen boundary conditions
effect the approximation to the whole space solution.

In this paper we address the problem of designing transparent boundary conditions (TBCs) 
for the 1D sine-Gordon equation using the 
so-called potential approach previously introduced in \cite{Matthias2008} (see, also the Refs.~\cite{Antoine1,Zhang} for further progress) 
and utilized in~\cite{Jambul1} for quantum graphs. 
Here we will adopt this approach for the sine-Gordon equation on a real line.
The motivation for the study of TBCs for the sine-Gordon equation comes from different practical important problems, 
such as tunable soliton transport in Josephson junctions \cite{Baron,BJJEPL}, energy transfer in DNA \cite{Yamosa,Yakushevich}, seismic waves and deformation 
propagation in tectonic plates \cite{Bykov1,Bykov2} and many others. 
In all these systems for certain cases one needs to achieve reflectionless propagation of waves and
particles to avoid different losses in charge, energy and signal transfer. 
This can be done by imposing TBCs 
for the governing wave equation and mapping these 
conditions on to physical characteristics of the system. Let us note that Zheng~\cite{Zheng07} presented a different, rather complicated approach for using TBCs for the sine-Gordon equation. Our approach is comparatively simple and more accessible for practitioners.

This paper is organized as follows. In the next section we give details of the procedure for the derivation 
of transparent boundary conditions for the sine-Gordon equation. 
Section~\ref{sec:discretization} presents a prescription for the discretization of these boundary conditions. 
Section~\ref{sec:NumericalExample} provides numerical results for modeling the propagation of sine-Gordon solitons with transparent boundary conditions and the 
explicit and energy conserving scheme of Fei and V\'azquez \cite{Fei94}. Finally, Section~\ref{sec:Conclusion} includes some concluding remarks.

\section{Transparent boundary conditions for the sine-Gordon equation}\label{sec:TBC}
Scattering of nonlinear waves at a given domain's boundary is a problem requiring to use an explicit solution of a 
wave equation describing these waves. 
However, the mathematical description of the absence of backscattering is a rather complicated task, since for nonlinear waves there is no S-matrix theory developed in quantum mechanics. 
Therefore, an effective solution for such problem can be to impose artificial boundary conditions 
for a wave equation, which describe the reflectionless transmission of the wave through the artificial boundary. 
TBCs for the evolution equations can be constructed by coupling the solutions of 
the initial value boundary 
problems (IVBPs) in the interior and exterior domains \cite{Arnold1998,Ehrhardt1999,Ehrhardt2001,Ehrhardt2002,Arnold2003,Jiang2004,Antoine2008,Ehrhardt2008,Sumichrast2009,Antoine2009,Ehrhardt2010,Klein2011,Arnold2012,Feshchenko2013,Antoine2014}.

Briefly, the general procedure for constructing 
transparent boundary conditions for a given PDE on a real 
line can be formulated as follows, cf.\ \cite{Antoine2008}
\begin{enumerate}\setlength{\itemsep}{0cm}
    \item Splitting the original wave equation into coupled equations, which are determined in 
    the interior and exterior domains on $\Omega^{\rm int}$, $\Omega^{\rm ext}$.
\item  Applying a Laplace transformation in time to the exterior problems on
$\Omega^{\rm ext}$. 
\item Solving the ordinary differential equations in the spatial 
variable $x$.  
\item Allowing only “outgoing” waves by selecting the asymptotically decaying solution as $x\to\pm\infty$.
\item Matching the Dirichlet and Neumann values at the artificial boundaries of the interirior domain. 
\item Applying (numerically) the inverse Laplace transformation.
\end{enumerate}

In this paper, following to the above procedure, 
we derive transparent boundary conditions for the \emph{sine-Gordon equation} (SGE) on a real line, which reads 
\begin{equation}\label{eq1}
  \partial_{x}^2 u-\partial_{t}^2 u-\sin{u}=0,\qquad x\in\mathbb{R},\;t>0,
\end{equation}
and is supplied with the following initial conditions:
\begin{equation}\label{eq:initial}
  u(x,0)=u_0(x), \qquad
   \partial_t u(x,0)=u_1(x).
\end{equation}

Let us note that Eq.~\eqref{eq1} admits a  \textit{soliton solution} in the form of a kink given by
\begin{equation}\label{eq:soliton}
   u(x,t) = 4\,\tan^{-1}\exp\biggl[\pm\frac{x-x_0-vt}{\sqrt{1-v^2}}\biggr],
\end{equation}
where $v$ denotes the (constant) velocity of the kink.
For completeness, we add (e.g., from \cite{drazin}) other solutions:
\begin{equation}\label{eq:breather}
   \text{Breather:}\quad u(x,t) = 4\,\tan^{-1}\biggl[\frac{\sqrt{1-v^2}}{v}\,
   \frac{\sin\bigl(v(t-t_0)\bigr)}{\cosh\bigl(\sqrt{1-v^2}(x-x_0)\bigr)}\biggr],
\end{equation}
\begin{equation}\label{eq:Antikink}
   \text{Kink-Antikink:}\quad u(x,t) = 4\,\tan^{-1}\biggl[\frac{v\cosh\bigl((x-x_0)/\sqrt{1-v^2}\bigr)}{\sinh\bigl(vt/\sqrt{1-v^2}\bigr)}
   \biggr].
\end{equation}

Furthermore, it is well-known \cite{Fei94} that the solutions to \eqref{eq1} conserve the \textit{total energy} (sum of kinetic, strain and potential energies)
\begin{equation}\label{eq:energy}
  E=\int_\mathbb{R}\biggl[
   \frac{1}{2}\bigl(\partial_t u(x,t)\bigr)^2+\frac{1}{2}\bigl(\partial_x u(x,t)\bigr)^2
   +G\bigr(u(x,t)\bigr)
   \biggr]\,dx=\mbox{const}.
\end{equation}
with the \textit{potential function} $G(u)=1-\cos{u}$, e.g.\ the kink~\eqref{eq:soliton} has the energy $8/\sqrt{1-v^2}$, and the momentum 
\begin{equation}\label{eq:momentum}
    P=-\int_\mathbb{R} (\partial_{t} u) (\partial_{t} u)\, dx = \frac{8v}{\sqrt{1-v^2}},
\end{equation}
cf.\ \cite{Fei94}. 
These invariants (or their discrete versions) 
can be used later to check the
usability of the considered numerical scheme.

Here we consider the propagation of a sine-Gordon soliton given by Eq.~\eqref{eq:soliton} on a finite interval $[0,L]$ 
and require its reflectionless transmission through the boundary of the interval, at $x=0$ and $x=L$ in terms of the boundary conditions for  Eq.~\eqref{eq1} .

For this purpose we apply the so-called \textit{potential approach}, 
which was earlier 
applied for the derivation of TBCs for the nonlinear Schr\"odinger equation~\cite{Antoine}.   
Within such an approach, one reduces the sine-Gordon equation~\eqref{eq1} 
into a linear PDE by introducing the following potential:
\begin{equation}\label{eq:potential}
   V(x,t)=-\frac{\sin{u}(x,t)}{u(x,t)}.
\end{equation}
It should be noted that this potential approach neglects 
here the dependency on the solution $u$ and considers it again at a later step.
Doing so, one can formally rewrite Eq.~\eqref{eq1} as the \textit{linear Klein-Gordon equation}
\begin{equation}\label{eq4}
  \partial_{x}^2 u-\partial_{t}^2 u+V(x,t) u=0,\qquad 0<x<L,\quad t>0.
\end{equation}

Next, introducing the new (unknown) function $v$, defined by the
$v(x,t)=e^{-\nu(x,t)}u(x,t)$, where
\begin{equation}\label{eq5}
     \nu(x,t)=\int_0^t\int_0^\tau V(x,s)\,dsd\tau,
\end{equation}
we obtain the following relations for the time and space derivatives of $u$:
\begin{equation*}
\begin{split} 
  \partial_t u&=e^{\nu} (\partial_t\nu+\partial_t)v,\\
  \partial_t^2 u&=e^{\nu}\bigl[ (\partial_t\nu)^2+2\partial_t\nu\cdot\partial_t+\partial_t^2+V \bigr]v,
\end{split}
\end{equation*}
and
\begin{equation*}
\begin{split} 
    \partial_x u&=e^{\nu}\left(\partial_x\nu+\partial_x\right)v,\\
    \partial_x^2 u&=e^{\nu}\bigl[ (\partial_x\nu)^2+2\partial_x\nu\cdot\partial_x+\partial_x^2+\partial_x^2\nu \bigr]v.
\end{split}
\end{equation*}
Then, the left hand side of Eq.~\eqref{eq4} for the new variable $v$ reads
\begin{equation}
  \mathcal{L}(x,t,\partial_x,\partial_t)v=\partial_x^2v
  -\partial_t^2v+A\partial_xv+(B-C)v-D\partial_tv,\label{eq8}
\end{equation}
where we have introduced the abbreviations
\begin{equation*} 
  A=2\partial_x\nu,\quad
  B=\partial_x^2\nu+(\partial_x\nu)^2,\quad
  C=(\partial_t\nu)^2,\quad
  D=2\partial_t\nu.
\end{equation*}
The operator $\mathcal{L}$ in Eq.~\eqref{eq8} can be formally factorized as
\begin{equation}
   \mathcal{L}=(\partial_x-\Lambda^-)(\partial_x+\Lambda^+)=\partial_x^2+(\Lambda^+-\Lambda^-)\partial_x+\Op(\partial_x\lambda^+)-\Lambda^-\Lambda^+.
   \label{eq10}
\end{equation}
Furthermore, we introduce the system of pseudo differential
operators~\cite{Taylor} and comparing with \eqref{eq8} leads to
\begin{equation}
\begin{split}\label{eq11}
  \Lambda^+-\Lambda^-&=A,\\
  \Op(\partial_x\lambda^+)-\Lambda^-\Lambda^+
  &=-\partial_t^2-D\partial_t+B-C,
\end{split}
\end{equation}
which yields the following system of equations on the symbol level:
\begin{equation}
\begin{split}\label{eq12}
  \lambda^+-\lambda^-&=a,\\
  \partial_x\lambda^+-\underset{\alpha=0}{\overset{+\infty}{\sum}}\frac{1}{\alpha!}\partial_\tau^\alpha\lambda^-\partial_t^\alpha\lambda^+
  &=-\tau^2-d\,\tau+b-c,
\end{split}
\end{equation}
where we have set $a=A$, $b=B$, $c=C$, $d=D$. The total symbol $\lambda^\pm$ of
the pseudo differential operator $\Lambda^\pm$ admits an asymptotic expansion
in inhomogeneous symbols as
\begin{equation}
   \lambda^\pm\sim\underset{j=0}{\overset{+\infty}{\sum}}\lambda_{1-j}^\pm.\label{eq13}
\end{equation}
If one considers only first order terms, then from the first equation
 one obtains $\lambda_1^-=\lambda_1^+$.
Accordingly, from the second equation of the system~\eqref{eq12} we have
\begin{equation}
  \lambda_1^+=\pm\tau.\label{eq14}
\end{equation}
For the potential $V(x,t)$, the Dirichlet-to-Neumann (DtN) formulation of the
TBC corresponds to the choice $\lambda_1^+=\tau$. For the zero order terms we
get
\begin{equation}
\begin{split}\label{eq15}
  \lambda_0^+-\lambda_0^-&=a,\\
   \partial_x\lambda_1^+-(\lambda_1^-\lambda_0^+ + \lambda_0^-\lambda_1^+)&=-d\,\tau,\quad \text{i.e.}\quad \partial_x\lambda_1^+-(\lambda_0^+ + \lambda_0^-)\tau=-d\,\tau.
\end{split}
\end{equation}
Using $\partial_x\lambda_1^+=0$ from \eqref{eq15} we have
\begin{equation}
   \lambda_0^+=\frac{a}{2}+\frac{d}{2}=\partial_x\nu+\partial_t\nu,\qquad\lambda_0^-=-\frac{a}{2}+\frac{d}{2}=-\partial_x\nu+\partial_t\nu.
   \label{eq16}
\end{equation}
For $j=2$ we have 
\begin{equation}
    \begin{split}
        \lambda_{-1}^+ - \lambda_{-1}^-&=0,\\
\partial_x\lambda_0^+ -(\lambda_1^-\lambda_{-1}^++\lambda_0^-\lambda_0^++\lambda_{-1}^-\lambda_1^++\partial_\tau\lambda_1^-\partial_t\lambda_0^++\partial_\tau\lambda_1^-\partial_t\lambda_{-1}^+)&=b-c,
    \end{split}
\end{equation}
since $\partial_\tau^\alpha\lambda_0^\pm=0,\,\partial_t^\alpha\lambda_1^\pm=0$,
$\alpha\in N$ and $\partial_\tau^\beta\lambda_1^\pm=0$,
$\beta\in\{2,3,4,\dots\}$. Now using \eqref{eq16} the second equation simplifies to
\begin{equation*}
   \begin{split}
\partial_t\lambda_{-1}^\pm+2\lambda_{-1}^\pm\tau
&=\partial_x\lambda_0^+ -\partial_t\lambda_0^+-\partial_x^2\nu\\
&=-\partial_{tt}^2\nu.
 \end{split}
\end{equation*}
From the last equation we obtain finally
\begin{equation}\label{eq:j2}
   \lambda_{-1}^-=\lambda_{-1}^+=-\int_0^tV(x,s)\cdot e^{-2\tau(t-s)}ds.
\end{equation}
This procedure can be continued for $j>2$. Now the DtN TBC applied to the function $v$ can be written as
\begin{equation} 
    \bigl(\partial_x\pm\Lambda^\pm\bigr)v=0,\label{eq18}
\end{equation}
or equivalently
\begin{equation}
   \bigl(\partial_x\pm\Lambda^\pm\bigr)e^{-\nu}u=0.\label{eq19}
\end{equation}

Following the Refs.~\cite{Antoine,Matthias2008}, 
we apply a ``cut-off'' (up
to $M-1$th term) in the expansion $\Lambda$ as
\begin{equation}\label{eq20}
    \Lambda_M^\pm=\Op\Biggl(\underset{j=0}{\overset{M-1}{\sum}}\lambda_{1-j}^\pm\Biggr).
\end{equation}

In this paper, we restrict ourselves to considering the expansion in Eq.~\eqref{eq20}
up to the third order approximation for 
transparent boundary conditions. Using \eqref{eq14}, \eqref{eq16} and \eqref{eq:j2}, we obtain
\begin{align*}\label{eq:lambdas}
\Lambda_1^{\pm}f&=\partial_tf,\\
\Lambda_2^{\pm}f&=\partial_tf\pm\partial_x\nu\cdot f+\partial_t\nu\cdot f,\\
\Lambda_3^{\pm}f&=\partial_tf\pm\partial_x\nu\cdot f+\partial_t\nu\cdot
f-e^{-2}\,\partial_t\nu\cdot f.
\end{align*}
 These results yield to the following TBCs. 

\smallskip
{\bf The first order approximation.}

For the left boundary (at $x=0$) we obtain the following expression:
\begin{equation}\label{eq:1approx0}
\partial_x u(0,t)=\Biggl[\partial_t u(x,t)+\biggl(\partial_x \nu(x,t) - \partial_t \nu(x,t)\biggr)\cdot u(x,t)\Biggr]_{x=0},
\end{equation}
Analogously, one can obtain the TBC for the right boundary (at $x=L$):
\begin{equation}\label{eq:1approxL}
\partial_x u(L,t)=\Biggl[-\partial_t u(x,t)+\biggl(\partial_x \nu(x,t) + \partial_t \nu(x,t)\biggr)\cdot u(x,t)\Biggr]_{x=L}.
\end{equation}

\smallskip
{\bf The second order approximation.}

For the left boundary (at $x=0$):
\begin{equation}\label{eq:2approx0}
\partial_x u(0,t)=\partial_t u(0,t),
\end{equation}
and for the right boundary (at $x=L$):
\begin{equation}\label{eq:2approxL}
\partial_x u(L,t)=-\partial_t u(L,t).
\end{equation}

\smallskip
{\bf The third order approximation.}

For the left boundary (at $x=0$):
\begin{equation}\label{eq:3approx0}
\partial_x u(0,t)=\Biggl[\partial_t u(x,t)-e^{-2}\cdot\partial_t \nu(x,t) \cdot u(x,t)\Biggr]_{x=0}.
\end{equation}
and for the right boundary (at $x=L$):
\begin{equation}\label{eq:3approxL}
\partial_x u(L,t)=\Biggl[-\partial_t u(x,t)+e^{-2}\cdot\partial_t \nu(x,t) \cdot u(x,t)\Biggr]_{x=L}.
\end{equation}

Eqs.~\eqref{eq:1approx0}-\eqref{eq:3approxL} represent approximations to the transparent boundary conditions for
the sine-Gordon equation \eqref{eq1}, which provide reflectionless
transport of the sine-Gordon solitons on a real line. It remains to implement these approximations in a numerical scheme,
which is a non-trivial task since these TBCs are nonlocal in time (of memory-type) with a singular kernel.

\section{Discretization sine-Gordon equation and transparent boundary conditions }\label{sec:discretization}

The efficient numerical implementation of the above transparent boundary conditions 
\eqref{eq:1approx0}--\eqref{eq:3approxL} is a non-trivial task and requires
using highly accurate and stable discretization schemes. We introduce the notation $k=\Delta t$, $h=\Delta x$, and $D_k^+$, $D_k^-$, $D_k^0$, $D_k^2=D_k^+D_k^-$ 
are the standard (forward, backward, centered, second order) difference quotients 
with step sizes in time $k$ or space $h$. Further, $\bigl(\cdot,\cdot\bigr)$ denotes the standard inner product on the real line, i.e.
\begin{equation}\label{eq:innerproduct}
    \bigl(u^n,v^n\bigr)=h \sum_{j\in\mathbb{Z}}u_j^n v_j^n,
\end{equation}
inducing the norm $\bigl|\bigl|u^n\bigr|\bigr|^2=  \bigl(u^n,u^n\bigr)$
and
the semi-norm 
\begin{equation}\label{eq:seminorm}
  \bigl|u^n\bigr|_1^2=\frac{1}{2}\bigl\|D_h^+u^n\bigr\|^2+\frac{1}{2}\bigl\|D_h^-u^n\bigr\|^2.
\end{equation}

\subsection{The standard discretization}

Let us recall that the standard discretization for the sine-Gordon equation~\eqref{eq1} uses central 
second difference quotients for approximating $\partial_t^2 u$, $\partial_x^2 u$ and reads
\begin{equation}\label{eq28standard}
\frac{u_j^{n+2}-2u_j^{n+1}+u_j^{n}}{\Delta t^2}
-\frac{u_{j+1}^{n+1}-2u_{j}^{n+1}+u_{j-1}^{n+1}}{\Delta x^2}
+\sin\bigr(u_{j}^{n+1}\bigl) =0,\quad j\in\mathbb{Z},\;n\ge0.
\end{equation}
i.e. in our notation
\begin{equation}\label{eq28standardnotation}
D_k^2 u_j^{n+1}
-D_h^2 u_{j}^{n+1}
+\sin\bigr(u_{j}^{n+1}\bigl) =0,\quad j\in\mathbb{Z},\;n\ge0.
\end{equation}
This leads to the following explicit scheme
\begin{equation}\label{eq29standard}
u_j^{n+2}=2\bigr(1-\gamma^2\bigl)u_j^{n+1}
               +\gamma^2\bigr(u_{j+1}^{n+1}+u_{j-1}^{n+1}\bigl)
              -\Delta t^2\sin\bigr(u_{j}^{n+1}\bigl) -u_{j}^{n},\quad j\in\mathbb{Z},\;n\ge0,
\end{equation}
where $\gamma=\Delta t/\Delta x$ denotes the hyperbolic mesh ratio.
For the starting step ($n=-1$) we use the central difference with the ghost value $u_j^{-1}$
\begin{equation*}
    \partial_tu(x_j,0)=u_1(x_j)=\frac{u_j^{1}-u_j^{-1}}{2\Delta t}+O(\Delta t^2)
\end{equation*}
and obtain from \eqref{eq29standard}
\begin{equation}\label{eq29standard_start}
u_j^1=\Delta t \,u_1(x_j) +\bigr(1-\gamma^2\bigl)u_j^{0}
               +\frac{\gamma^2}{2}\bigr(u_{j+1}^{0}+u_{j-1}^{0}\bigl)
              -\frac{\Delta t^2}{2}\sin\bigr(u_{j}^{0}\bigl) ,
\end{equation}
with  the initial data $u_{j}^{0}=u_0(x_j)$, $j=0,1,\dots J$.

For checking the discrete energy conservation (and thus the stability and the suitability to model the long time behavior of the solution) we multiply \eqref{eq28standardnotation} with the central difference quotient $D_k^0 u_j^{n+1}=(u_j^{n+2}-u_j^n)/(2k)$ and obtain
\begin{equation}\label{eq28standardnotation_energy}
\frac{1}{2}D_k^-\bigl(D_k^+ u_j^{n+1}\bigr)^2
-\bigl(D_k^0 u_j^{n+1}\bigr)\bigl(D_h^-D_h^+ u_j^{n+1}\bigr)
+\bigl(D_k^0 u_j^{n+1}\bigr)\sin\bigr(u_j^{n+1}\bigl) =0,\quad j\in\mathbb{Z},\;n\ge0.
\end{equation}
Next, summing over $j\in\mathbb{Z}$ and summation by parts yields
\begin{equation}\label{eq28standardnotation_energy2}
D_k^-\sum_{j\in\mathbb{Z}}\frac{1}{2}\bigl(D_k^+ u_j^{n+1}\bigr)^2
+\sum_{j\in\mathbb{Z}}\bigl(D_k^0 D_h^+ u_j^{n+1}\bigr)\bigl(D_h^+ u_{j}^{n+1}\bigr)
+\sum_{j\in\mathbb{Z}}\bigl(D_k^0 u_j^{n+1}\bigr)\sin\bigr(u_{j}^{n+1}\bigl) =0,\quad n\ge0.
\end{equation}

\subsection{An explicit energy conserving scheme}

The third term in \eqref{eq28standardnotation_energy2} arising from the standard discretization prevents a proper energy conservation and for this reason we modify the sine term in the scheme:
\begin{equation}\label{eq28schemeA}
\frac{u_j^{n+2}-2u_j^{n+1}+u_j^{n}}{\Delta t^2}
-\frac{u_{j+1}^{n+1}-2u_{j}^{n+1}+u_{j-1}^{n+1}}{\Delta x^2}
= \frac{\cos\bigr(u_{j}^{n+2}\bigl)-\cos\bigr(u_{j}^{n}\bigl)}{u_{j}^{n+2}-u_{j}^{n}},\quad j\in\mathbb{Z},\;n\ge0.
\end{equation}

The right hand side of Eq.~\eqref{eq28schemeA} is a second order approximation to 
$-\sin\bigr(u_{j}^{n+1}\bigl)$ which explains the consistency to \eqref{eq1}.

It can be shown using the same steps as in \eqref{eq28standardnotation_energy}, 
\eqref{eq28standardnotation_energy2} that this implicit  scheme~\eqref{eq28schemeA}
satisfies on $j\in\mathbb{Z}$ a discrete analogue of the energy conservation \eqref{eq:energy}, cf.\ \cite{Fei94}
\begin{equation}\label{eq:energy0}
E_0^{n+1}=
h\sum_{j\in\mathbb{Z}}\biggl[ \frac{1}{2}
\bigl(D_k^+u_j^n\bigr)^2
+\frac{1}{2}\bigl(D_h^+u_j^{n+1}\bigr)\bigl(D_h^+u_j^n\bigr)
+\frac{G(u_{j}^{n+1})+G(u_{j}^{n})}{2}\biggr]=\mbox{const}.
\end{equation}

Additionally, we modify the temporal discretization in  Eq.~\eqref{eq28schemeA} to obtain an efficient explicit scheme:
\begin{equation}\label{eq28schemeB}
\frac{u_j^{n+3}-\bigl(u_j^{n+2}+u_j^{n+1}\bigr)+u_j^{n}}{2\Delta t^2}
-\frac{u_{j+1}^{n+2}-2u_{j}^{n+2}+u_{j-1}^{n+2}}{\Delta x^2}
-\frac{u_{j+1}^{n+1}-2u_{j}^{n+1}+u_{j-1}^{n+1}}{\Delta x^2}
= \frac{\cos\bigr(u_{j}^{n+2}\bigl)-\cos\bigr(u_{j}^{n+1}\bigl)}{u_{j}^{n+2}-u_{j}^{n+1}},
\end{equation}
$j\in\mathbb{Z}$, $n\ge0$,
which was proposed by Fei and Vazquez \cite{Fei94} as `Scheme~1 (S1)' and reads in our notation
\begin{equation}\label{eq28schemeBnotation}
D_k^2 u_j^{n+3/2} - D_h^2 u_{j}^{n+3/2}
+ \frac{G\bigr(u_{j}^{n+2}\bigl)-G\bigr(u_{j}^{n+1}\bigl)}{u_{j}^{n+2}-u_{j}^{n+1}}=0,\quad j\in\mathbb{Z},\;n\ge0.
\end{equation}
Here, we have introduced the arithmetic averaging $u_j^{n+3/2}= \bigl(u_j^{n+2} +u_j^{n+1}\bigr)/2$.
The solution $u_j^{n+3}$ can be computed explicitly from the difference equation \eqref{eq28schemeB}, once the starting values
$u_j^0$, $u_j^1$, $u_j^2$ are available.
Also, the scheme \eqref{eq28schemeB} is second order in time and space and fulfills for $j\in\mathbb{Z}$ the discrete energy conservation, cf.\ \cite{Fei94}
\begin{equation}\label{eq:energy1}
E_1^{n+1}=h\sum_{j\in\mathbb{Z}}\biggl[
\frac{1}{2}\bigl(D_k^+u_{j}^{n+1}\bigr)\bigl(D_k^-u_{j}^{n+1}\bigr)
+ \frac{1}{2}\bigl(D_h^+u_{j}^{n+1}\bigr)^2
+G(u_{j}^{n+1})\biggr]=\text{const}.
\end{equation}
Let us note that Vu-Quoc and Li \cite{Vu-Quoc93, Li95} 
investigated the construction of energy conserving finite difference schemes for nonlinear Klein-Gordon equations in a general setting.

\subsection{Implementation of the TBC}

In this subsection we present our numerical method for finding values of the wave function at transparent boundaries. Here we give prescription only for the TBC of the third approximation \eqref{eq:3approxL} at $x=L$. We note that for the first and second approximations the implementation can be done analogously. Thus, denoting $g=u_J^n$,
in each time step one needs to find zero of the following function
\begin{equation}\label{eq:funczero}
f(g)=\gamma\left(g-u_{J-1}^n\right) + \left(g-u_J^{n-1}\right) -\Delta t\,e^{-2}\,\partial_t\nu^n(g)\cdot g.
\end{equation}
The function zeros can be found using the Newton-Raphson method, for which the derivative of the function is required:
\begin{equation}\label{eq:derfunczero}
f'(g)=\gamma + 1 -\Delta t\,e^{-2}\,\biggl(\bigl[\partial_t\nu^n(g)\bigr]'\cdot g+\partial_t\nu^n(g)\biggr).
\end{equation}

We discretize the double integral function $\nu(x,t)\approx \nu^n(x)$ ginven by \eqref{eq5} using the trapezoidal rule in the following way
\begin{align*}
   \nu^n(x)&=\nu^{n-1}(x)+\int_{t_{n-1}}^{t_n} \int_0^\tau V(x,s)\,ds d\tau
   = \nu^{n-1}(x) + \frac{\Delta t}{2} \left(\int_0^{t_{n-1}} V(x,s)\,ds
      +\int_0^{t_{n}} V(x,s)\,ds\right)\\
  &= \nu^{n-1}(x) + \frac{\Delta t}{2}\left[\frac{\Delta t}{2}\left(V^0(x)+2\sum_{k=1}^{n-2} V^k(x)+V^{n-1}(x)\right)
  +\frac{\Delta t}{2} \left(V^0(x)+2\sum_{k=1}^{n-1} V^k(x)+V^{n}(x)\right)\right]\\
 &= \nu^{n-1}(x) + \frac{\Delta t^2}{4}\left(2V^0(x)
    + 4\sum_{k=1}^{n-2} V^k(x) + 3V^{n-1}(x) + V^{n}(x)\right),\quad n\ge2,
\end{align*}
where $\nu^{0}(x)=0$ and $\nu^{1}(x)=\dfrac{\Delta
t^2}{4}\left(V^0(x)+V^1(x)\right)$.

In the same way one can discretize 
$$
\partial_t\nu(x,t)=\int\limits_0^t V(x,s)\,ds
$$
using the same trapezoidal rule as
$$
\partial_t\nu(x,t)\approx \partial_t\nu^n(x)=\frac{\Delta t}{2}\Bigl(V^0(x)+2\sum_{k=1}^{n-1}{V^k(x)}+V^{n}(x)\Bigr).
$$

Similarly, $\partial_x\nu(x,t)$ (which is needed in the first order approximation) can be approximated as follows
$$
\partial_x\nu(x,t)=\int\limits_0^t\int\limits_0^\xi{\partial_xV(x,s)\,dsd\xi} \approx \partial_x\nu^n(x),
$$
where
$$
\partial_x\nu^n(x)=\partial_x\nu^{n-1}(x)+\frac{\Delta t^2}{4}\Biggl(2\partial_xV^0(x)
+4\sum_{k=1}^{n-2}{\partial_xV^k(x)}+3\partial_xV^{n-1}(x)+\partial_xV^{n}(x)\Biggr)
$$
with $\partial_xV^k(x_j)=\partial_g\Bigl(-\dfrac{\sin
g}{g}\Bigr)\biggr|_{g=u_j^k}\cdot\dfrac{u_j^k-u_{j-1}^k}{\Delta x}$.

For the derivative in \eqref{eq:derfunczero} one can use the following approximation
\begin{equation*}
\bigl[\partial_t\nu^n(g)\bigr]' = \frac{\Delta t}{4}F'(g),\,\,n\ge1,
\end{equation*}
with $\bigl[\partial_t\nu^0(g)\bigr]'=0$.

In the next section, where we present a numerical example, we use this prescription in our numerical calculations of TBCs.

\subsection{Stability of the overall scheme}
It remains to check the stability of the scheme \eqref{eq28schemeBnotation} on a bounded grid 
supplied with our discretized TBC at $x=0$, $x=L$ (i.e.\ $j=0$, $j=J$). Thus
we have to consider the inner product $(\cdot,\cdot)$ on a finite range $j=0,1,\dots,J$ 
\begin{equation}\label{eq:innerJ}
    \bigl(u^n,v^n\bigr)_J=h \sum_{j=1}^{J-1} u_j^n v_j^n,
\end{equation}
with the corresponding induced norms and semi-norms as in \eqref{eq:seminorm}.
Next, we multiply \eqref{eq28schemeBnotation} by $D_{k/2}^0 u_j^{n+3/2}=D_k^+ u_j^{n+1}=(u_{j}^{n+2}-u_{j}^{n+1})/k$ and take the inner product~\eqref{eq:innerJ} 
\begin{equation}\label{eq28schemeBnotationmult}
h\sum_{j=1}^{J-1} D_k^+ u_j^{n+1} D_k^2 u_j^{n+3/2} 
- h\sum_{j=1}^{J-1} D_k^+ u_j^{n+1}D_h^2 u_{j}^{n+3/2}
+ \frac{h}{k}\sum_{j=1}^{J-1}\Bigl(G\bigr(u_{j}^{n+2}\bigl)-G\bigr(u_{j}^{n+1}\bigl)\Bigr)=0,\quad j\in\mathbb{Z},\;n\ge0.
\end{equation}
An easy calculation proves the following identity for the first term, cf.\ \eqref{eq:energy1}
\begin{equation}\label{eq28:identity}
   D_k^+ u_j^{n+1} D_k^2 u_j^{n+3/2} 
   = D_k^+ \frac{1}{2}\bigl(D_k^+u_j^{n+1}\bigr)\bigl(D_k^-u_j^{n+1}\bigr).
\end{equation} 
Then we apply the summation by parts rule for two grid functions $f_j$, $g_j$ on a finite index range
\begin{equation}\label{eq:summationbypartsrule}
h\sum_{j=1}^{J-1}g_j D_h^-f_j
=-h\sum_{j=0}^{J-1}f_j D_h^+g_j
+f_{J-1}g_J- f_0g_0
\end{equation}
and obtain choosing $g_j=D_k^+ u_j^{n+1}$, $f_j=D_h^+ u_j^{n+3/2} $
\begin{multline}\label{eq28standardnotation_energy2J}
D_k^+ h\sum_{j=1}^{J-1}\frac{1}{2}\bigl(D_k^+u_j^{n+1}\bigr)\bigl(D_k^-u_j^{n+1}\bigr)
+h\sum_{j=0}^{J-1}\bigl(D_h^+ u_j^{n+3/2}\bigr)D_h^+\bigl(D_k^+ u_j^{n+1}\bigr)
+D_k^+ h
\sum_{j=1}^{J-1}G\bigr(u_{j}^{n+1}\bigl)
\\
=
 \bigl(D_h^- u_J^{n+3/2}\bigr) \bigl(D_k^+ u_{J}^{n+1}\bigr) 
- \bigl(D_h^+ u_0^{n+3/2}\bigr)   \bigl(D_k^+ u_{0}^{n+1} \bigr)
,\quad n\ge0.
\end{multline}
Another elementary algebraic calculation shows for the second term, cf.\ \eqref{eq:energy1}
\begin{equation}\label{eq28standardnotation_energy2Jterm}
\bigl(D_h^+ u_j^{n+3/2}\bigr)D_h^+\bigl(D_k^+ u_j^{n+1}\bigr)
=D_k^+\frac{1}{2}\bigl(D_h^+u_{j}^{n+1}\bigr)^2, 
\end{equation}
i.e.\ we obtain
\begin{multline}\label{eq28standardnotation_energy2JJ}
D_k^+ h\sum_{j=1}^{J-1}\biggl[\frac{1}{2}\bigl(D_k^+u_j^{n+1}\bigr)\bigl(D_k^-u_j^{n+1}\bigr)
+\frac{1}{2}\bigl(D_h^+u_{j}^{n+1}\bigr)^2 +G(u_{j}^{n+1})\biggr]
\\
= \bigl(D_h^- u_J^{n+3/2}\bigr) \bigl(D_k^+ u_{J}^{n+1}\bigr) 
- \bigl(D_h^+ u_0^{n+3/2}\bigr)   \bigl(D_k^+ u_{0}^{n+1} \bigr),\quad n\ge0.
\end{multline}
The left hand side of \eqref{eq28standardnotation_energy2JJ}
is exactly the discrete time derivative $D_k^+ E_1^{n+1}$ of the discrete energy 
defined in \eqref{eq:energy1}, i.e.\ for a stable overall scheme
one has to check finally (possibly only numerically for nonstandard boundary conditions like the TBCs) if the right hand side of \eqref{eq28standardnotation_energy2JJ} (the boundary terms) are negative, such that the discrete energy will decay on the finite interval.
E.g. in the simple cases of Dirichlet boundary conditions we have $D_k^+ u_{0}^{n+1}=0$, $D_k^+ u_{J}^{n+1}=0$ and for homogeneous Neumann boundary conditions
the discrete normal derivatives
$D_h^+ u_{0}^{n+3/2}$, $D_h^+ u_{J}^{n+3/2}$ vanish, i.e.,
for these standard boundary conditions the right hand side of \eqref{eq28standardnotation_energy2JJ} is zero, the discrete energy $E_1^{n+1}$ is conserved and thus the overall scheme is stable.

Analogously one can check the sign of the boundary terms 
in \eqref{eq28standardnotation_energy2JJ}
 for the TBCs,
e.g.\ the second order approximation \eqref{eq:2approx0}, \eqref{eq:2approxL} is discretized as follows
\begin{equation}\label{eq:2approx0Ld}
D_h^+ u_{0}^{n+3/2}=D_k^+ u_{0}^{n+1},\qquad
D_h^+ u_{J}^{n+3/2}=-D_k^+ u_{J}^{n+1}
\end{equation}
and thus the right hand side of \eqref{eq28standardnotation_energy2JJ} is negative, i.e.\ the discrete energy $E_1^{n+1}$ decays and thus the overall scheme is stable.

\begin{figure}[t!]
\includegraphics[width=10cm]{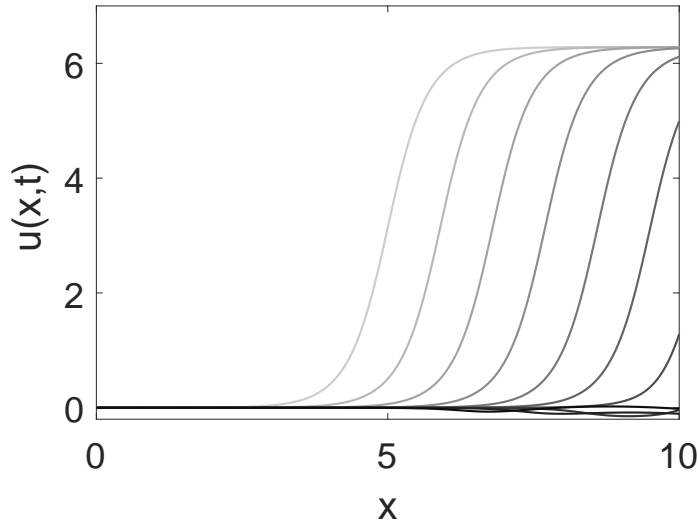}
\caption{Evolution of a kink~\eqref{eqkink} simulated with finite difference scheme given by~\eqref{eq28schemeA}. 
This plot is obtained for TBC of the third order approximation \eqref{eq:3approxL}.}\label{wpd}
\end{figure}

\begin{figure}[t!]
\includegraphics[width=10cm]{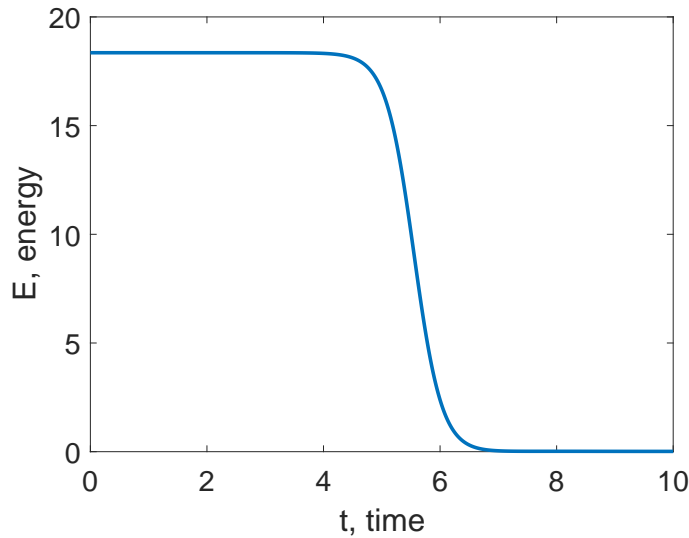}
\caption{Time evolution of the discrete 
kink energy \eqref{eq:energy1} in the interior domain $[0,L]$.}\label{ener1}
\end{figure}

\section{Numerical Example}\label{sec:NumericalExample}

Now we solve the sine-Gordon equation~\eqref{eq1} on the finite interval $[0,L]$
imposing TBC at the right boundary (i.e.\ at $x=L$). As the initial conditions we choose a kink-soliton at $t=0$ 
\begin{equation}\label{eqkink}
   u(x,0) = 4\,\tan^{-1}\exp\biggl[\frac{x-x_0}{\sqrt{1-v^2}}\biggr],
\end{equation}
and its time derivative (at $t=0$):
\begin{equation}\label{deqkink}
   \partial_t u(x,0) = -2\,\frac{v}{\sqrt{1-v^2}}\sech\biggl[\frac{x-x_0}{\sqrt{1-v^2}}\biggr].
\end{equation}
We apply the explicit energy conserving scheme~\eqref{eq28schemeB} using the following parameters set: 
space interval $L=10$, space discretization step $\Delta x=0.02$, 
time step $\Delta t=0.0002$ and velocity of the kink $v=0.9$ and its center position $x_0=5$.

In Fig.~\ref{wpd} the evolution of a kink on the space interval $[0,10]$ with TBCs of the third order approximation is presented.
The corresponding energy evolution is plotted in Fig.~\ref{ener1} 
using its discrete analogue given by \eqref{eq:energy1}.
From this plot one can observe that the total energy vanishes with the transition of the wave function 
through the artificial boundary, which implies that a kink completely leaves the interval $[0,L]$ without reflection at the boundary.  
For further analyses we consider three time intervals of the dynamics: $[0,3]$ -- period, during which no influence of TBCs is observed; $(3,7.8]$ -- within this period the kink passes through the artificial boundary; $(7.8,10]$ -- time left after the kink's transition. 

\begin{figure}[t!]
\includegraphics[width=8.5cm]{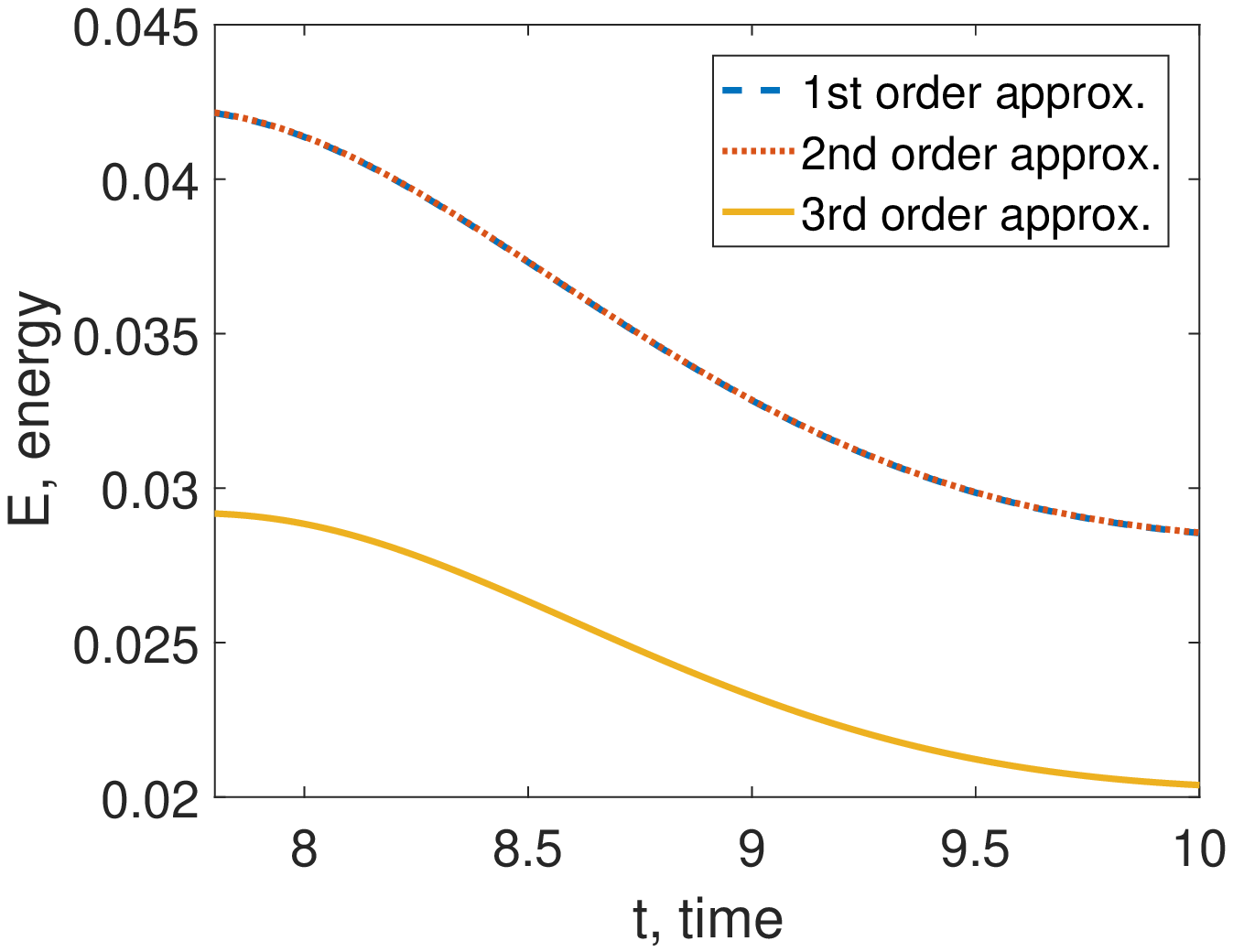}
\includegraphics[width=8.5cm]{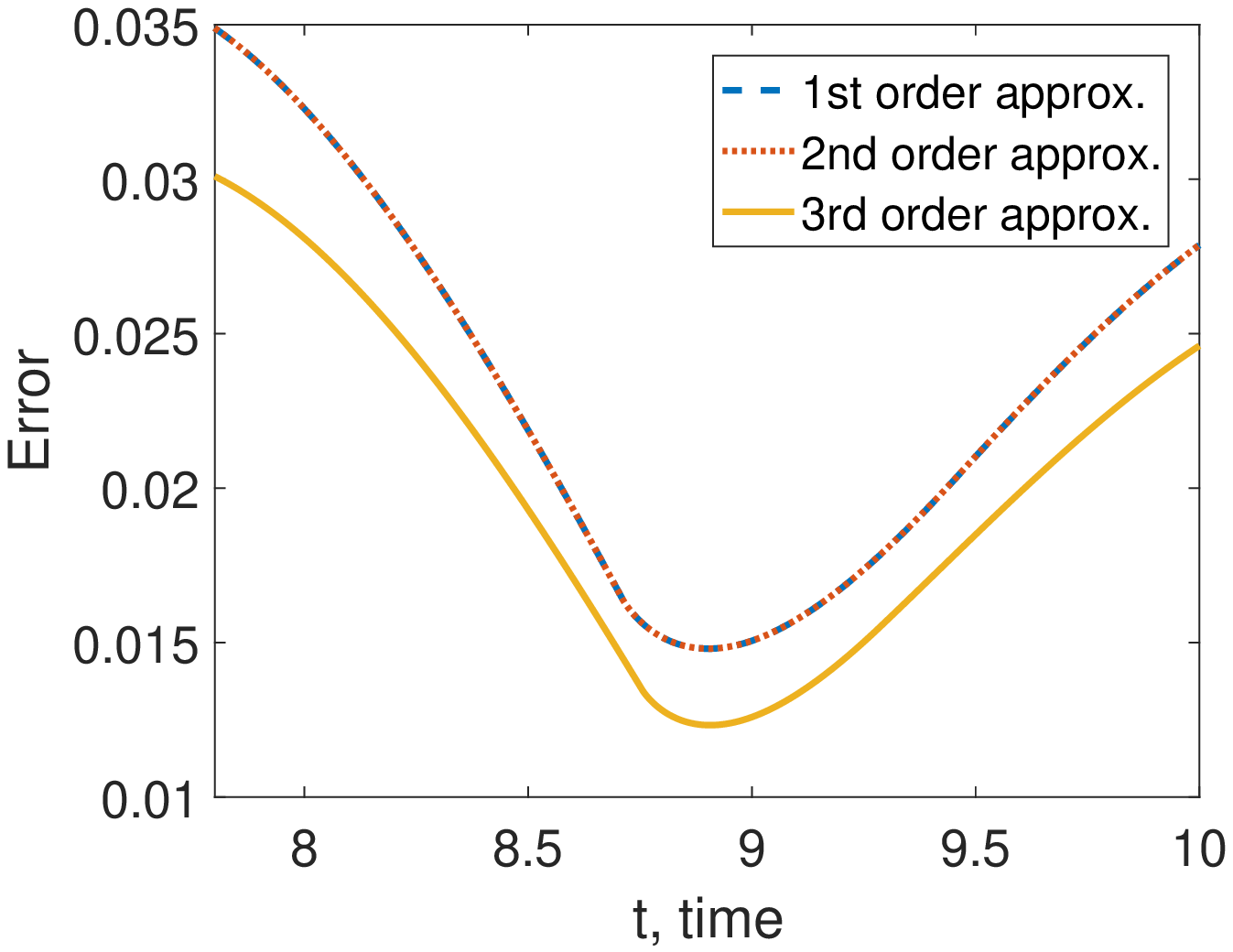}
\caption{Comparison of total energies (left panel) and total error (right panel) calculated for the first, second and third order approximations within the last considered time period.}\label{fig:compener}
\end{figure}

As the energy must vanish after the kink passes the artificial boundary (i.e. for the third time interval), in Fig.~\ref{fig:compener} in the left panel it is shown that the energy decreases with higher order TBCs. This can also be checked by computing the error defined as 
\begin{equation}\label{eq:error1}
   \text{ER}(n\Delta t)=\frac{1}{J-1}\sum_{j=1}^{J-1}{\bigl|u_j^n\bigr|}.
\end{equation}.

In Fig.~\ref{fig:compener} in the right panel one can observe that the error decreases with higher order TBCs.

In our numerical investigations we further check the variations in 
the  discrete momentum \eqref{eq:momentum}
 \begin{equation}\label{eq:discretemomentum}
    P^{n+1}=-h\sum_{j\in\mathbb{Z}}(D^0_ku_j^{n+1})(D^0_hu_j^{n})
    =-h\sum_{j\in\mathbb{Z}}\frac{u_j^{n+2}-u_j^{n}}{2k} \frac{u_{j+1}^{n+1}-u_{j-1}^{n+1}}{2h}.
\end{equation}

\begin{figure}[t!]
\includegraphics[width=8.5cm]{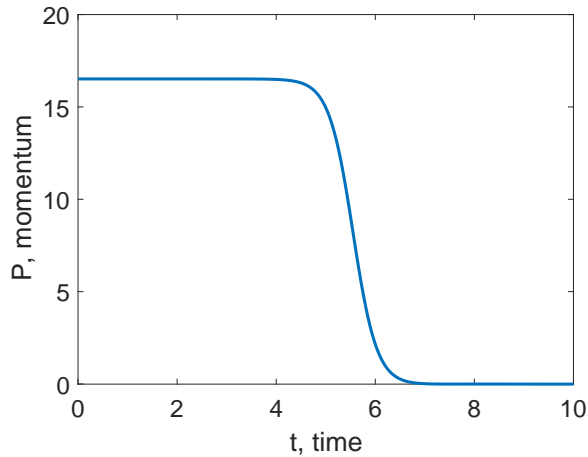}
\caption{Time evolution of the discrete momentum \eqref{eq:discretemomentum} in $[0,L]$}\label{fig:momentum}
\end{figure}

Time evolution of the discrete momentum is shown in Fig.~\ref{fig:momentum}. This plot demonstrates the same behaviour as that for the energy, which implies reflectionless transmission of a kink-soliton through the artificial boundary.

\section{Conclusions}\label{sec:Conclusion} 
We have derived explicit transparent boundary conditions for the 
sine-Gordon equation on a real line.
The so-called potential approach is used for reducing the sine-Gordon equation 
to the linear Klein-Gordon equation. 
An effective and stable discretization for transparent boundary 
conditions is proposed and implemented to model
the reflectionless propagation of sine-Gordon solitons on a line. 
A stability analysis and error estimates for the numerical method are provided.  
The above results can be directly used for modeling the transport of sine-Gordon solitons 
in a broad variety of physical systems and processes, such as Josephson junctions, deformation propagation in solids, energy transport in DNA and seismic waves in tectonic plates. Although the above treatment 
deals with the kink solitons, similar approach can be applied for other travelling wave solutions of sine-Gordon equation. 

In future work we will extend our approach to two dimensions.
Also, we will consider discrete TBCs that are designed directly for the considered numerical scheme.
Finally we will transfer our TBCs for sine-Gordon equations on metric graphs, 
that are needed at the branching points, as it was done for the nonlinear Schr\"odinger equation in \cite{Jambul1}.



\end{document}